\documentstyle[twocolumn,pre,aps]{revtex}

\catcode`\@=11

\def\maketitle2{\par % Uses \twocolumn[\@maketitle2].
\begingroup
\let\cite\@bylinecite
\def\thefootnote{\fnsymbol{footnote}}%
\twocolumn[\@maketitle2\vskip2pc]%
\thispagestyle{plain}\@thanks
\endgroup
\def\thefootnote{\arabic{footnote}}%
\setcounter{footnote}{0}%
\let\maketitle2\relax \let\@maketitle2\relax
\let\@thanks\relax \let\@authoraddress\relax \let\@title\relax
\let\@date\relax \let\thanks\relax \let\@abstract\relax
\let\@pacs\relax}

\def\abstract#1{\gdef\@abstract{{\par % Store abstract text.
\bgroup
\ifdim\prevdepth=-1000pt \prevdepth0pt\fi
\hsize\columnwidth
\dimen0=-\prevdepth \advance\dimen0 by17.5pt \nointerlineskip
\small\vrule width 0pt height\dimen0 \relax}{~~}#1\egroup}}

\def\pacs#1{\gdef\@pacs{{\par % Store PACS numbers as \@pacs.
\bgroup
\hsize\columnwidth \parindent0pt
\ifdim\prevdepth=-1000pt \prevdepth0pt\fi
\dimen0=-\prevdepth \advance\dimen0 by20pt\nointerlineskip
\egroup} PACS numbers:~#1}}

\def\@maketitle2{% Puts \@abstract and \@pacs in a {list}.
\@preprint
\@title
\ifdim\prevdepth=-1000pt \prevdepth0pt\fi
\@authoraddress
\@date
\begin{list}{}{\leftmargin=0.10753\textwidth \rightmargin=\leftmargin
\itemsep=1pc\partopsep=-1pc}
\item\@abstract
\item\@pacs
\end{list}
}

\catcode`\@=12

\begin{document}

\preprint{INP preprint, April 1998}

\draft

\title{Comment on fractality of quantum mechanical energy spectra}

\author{Andrzej Z. G\'orski\thanks{Address: Institute of Nuclear
Physics, Radzikowskiego 152, 31--342 Krak\'ow, Poland, e--mail:
gorski@alf.ifj.edu.pl}}
\address{Institute of Nuclear Physics, Cracow, Poland}

\date{\today}

\abstract
{\small{
 The fractal properties of the energy spectra of
quantum systems are discussed in connection with the paper by
S\'aiz and Mart\'inez [Phys. Rev. E {\bf 54}, 2431 (1996)].
It is shown that for discrete
energy levels the Hausdorff--Besicovitch dimension is zero
and differs from the Renyi scaling exponents computed by
the standard box counting algorithm.
 The Renyi exponents for the inverse power series data sets ($x_n =
{1\over n^a}, \ n=1,2,\ldots$) are computed analytically
and they are shown to be $d_0 = {1\over 1+a}$
and, as a consequence, $d_0 = 1/3$ for the Balmer formula.
}}

\pacs{05.45.+b, 32.30.-r, 24.60.Lz}

\maketitle2
\narrowtext

 In last years there is an increasing interest in searching for fractal
structures in physics \cite{Mandelbrot}
and in looking for fractal signatures of chaos at the quantum mechanical
(QM) level \cite{StasEPSAPS,Saiz,StasAPP}, as well.
 In fact, a simple QM system  with fractal energy spectrum has been
found long time ago \cite{Hofstadter}.
Similar structures have been found in recent years in other systems
that are of great practical importance: the quasiperiodic semiconductor
microstructures, the quantum Hall effect
and the Anderson localization
\cite{GeiselPHYSICA,localization}.
 In this paper I would like to comment on possible fractality of QM
spectra and I will discuss Renyi exponents \cite{RenyiA,RenyiB}
for some special types of point spectra generated by
simple probability distributions, as well as for Balmer like
energy levels computed in \cite{Saiz}
\begin{equation}
E_n \sim \left( {1\over m^2} - {1\over n^2} \right)
\ , \quad  n>m \ , \quad  m=1,2,3,4,5 \ .
\label{Balmer}
\end{equation}

 Typical QM systems have either discrete point spectra (localized
states, like for the harmonic oscillator), continuous spectra
(extended states, like for the free particle) or both
(localized and extended states above some threshold energy, like for the
hydrogen atom).
In addition, the models have been found that have neither extended
nor localized states and their energy spectra have been shown
to be fractal \cite{Hofstadter,GeiselPHYSICA,localization}.
These are models that describe infinite crystals
(periodic or almost periodic).
 In this paper I will limit discussion to the case of discrete
spectra.

 For the discrete spectra it has been shown that their nearest neighbor
spacing (NNS) of energy levels has Poisson or Wigner
probability distribution (for corresponding chaotic Hamiltonians)
\cite{Wigner,Haake}.
Such spectra cannot lead to any fractal structure as any
reasonable probability distribution cannot give the Cantor like
structures (in particular, the Poisson, Wigner or Thomas--Porter
distributions are regular enough).
On the other hand, the energy spectra with finite number of accumulation
points cannot be generated by a reasonable (regular) probability
distribution.
However, using the correct definition of the fractal
(Hausdorff--Besicovitch) dimension \cite{GrassbergerHausDim,Falconer}
one can easily show that for any set with finite number of accumulation
points the Hausdorff--Besicovitch dimension is zero, $d_H = 0$.
Hence, for any nuclear or molecular discrete energy spectra we should
get their fractal dimension equal to zero.

 In practical (numerical) calculations of the fractal
dimension the box counting algorithm and the Renyi exponents
(improperly called ``dimensions'') are used \cite{Molteno}.
The whole interval is divided into $N$ equal subintervals
(``boxes'') with $n_i(N)$ ($i=1,2,\ldots, N$) data points in each box.
Defining the measure $p_i(N) = n_i(N)/n_{tot}$, where
$n_{tot} = \sum_i n_i(N)$ is the
total number of data points, we have the following definition of the
Renyi exponents
\begin{equation}
d_q = {-1\over q-1} \lim_{N\to\infty} {\ln \sum_i p_i^q(N)
\over \ln N}  \ .
\label{RenyiDEF}
\end{equation}
Here, for $q=0$ (the capacity ``dimension'') we assume summation over
non--empty boxes only, and (\ref{RenyiDEF}) becomes equivalent to
\begin{equation}
d_0 = \lim_{N\to\infty} { \ln M(N) \over \ln N }
\ ,
\label{RenyZeroDEF}
\end{equation}
where $M(N)$ is the total number of non--empty boxes.
For $q=1$ (the information ``dimension'') the de l'Hospital rule is
applied to the formula (\ref{RenyiDEF}) and one gets
\begin{equation}
d_1 = - \lim_{N\to\infty} {\sum_i p_i(N) \ln p_i(N)
\over \ln N}  \ .
\label{RenyOneDEF}
\end{equation}

 Once again, it is important to remember that the Renyi exponents are
{\em not} the (fractal) dimensions and for some ``pathological'' sets
they may differ from the Hausdorff--Besicovitch dimension
\cite{GrassbergerHausDim,Falconer}.
In particular, this happens for the inverse power like
discrete sets, as will be shown below.
However, even for such pathological sets one can use the Renyi
exponents just to describe the scaling properties of the data.
With this remark in mind we prove the following \\
{\em Theorem:}
For the inverse power series $x_n = 1/n^a$, where $n=1,2,3,\ldots$
and $a>0$ the Renyi exponent $d_0$ is
\begin{equation}
d_0 = {1\over 1 + a}
\ .
\label{ZeroExp}
\end{equation}

{\em Proof:}
Let us denote by $n_{sngl}$ the number of points that are single
in their boxes. We have $M(N) \ge n_{sngl}$ and,
from the condition $x_n - x_{n+1} \ge 1/N$
(distance between such points is greater than the box size $=1/N$),
we get $n_{sngl} = a^{1\over 1+a} N^{1\over 1+a}$.
In fact, in the last equation there should be taken integer
part of the right hand side but this is unimportant as we are
interested in the limit $N\to\infty$.
Using the definition \ref{RenyiDEF} with $q=0$ one gets
asymptotically for $N\to\infty$ the
lower limit $d_0 \ge (a+1)^{-1}$. \\
 To get the lower limit we took into account points in the
interval $[x_{sngl}, 1]$, $x_{sngl} = x_{n_{sngl}} = 1/n_{sngl}^a$.
To obtain an upper limit we have to take into account the remaining
boxes as well. First, let us notice that the first box,
{\em i.e.} the interval $[0, 1/N]$ contains infinite number of
points for any $N$, as zero is the accumulation point, and
the interval $[1/N, x_{sngl}]$ contains at most
$(x_{sngl} - 1/N) / (1/N)$ non--empty boxes. This gives
the following upper limit for $M(N)$
\[
M(N) \le a^{1\over 1+a} \left( 1 + {1\over a} \right)
N^{1\over 1+a}   \ ,
\]
and, as a consequence, we have
\[
d_0 \ \le \ \lim_{N\to\infty}
{ \ln \left[ a^{1\over a+1} \left( 1 + {1\over a} \right) N^{1\over a+1}
\right]  \over  \ln N } \ = \ {1\over a+1}
\ .
\]
Hence, we have proven the eq. (\ref{ZeroExp}).

 This result gives us the scaling exponent $d_0 = 1/2$ for
the harmonic series (as was stated in \cite{Saiz}),
while for the Balmer like series ($a=2$)  we have
$d_0 = 1/3 \simeq 0.33$, in contrast to the result presented
in \cite{Saiz}, where the value 0.61 was obtained numerically.
 Of course, neither the Hausdorff--Besicovitch dimension nor the
Reny exponent be can changed by the overall shift of the whole set
(due to the $1/m^2$ term in (\ref{Balmer})), or by rejection of any
finite number of the data points ($n>m$), or even by superposition
of any finite number of such sets (as we have in (\ref{RenyZeroDEF})
logarithm in the numerator). In fact, superposition of two fractal sets
has the fractal dimension equal to the maximum fractal dimension of
the constituent sets \cite{Falconer}.
Hence, our result cannot be changed by any simple composition
of Balmer like sets.

 I have done some numerical checks of the analytical result
(\ref{ZeroExp}) with the box counting algorithm and both methods
are consistent. In particular, for $a=2$ and $1000$ data points
the box counting method gives: $d_0 \simeq 0.321$,
where 6 points fit to the straight line with $\chi^2 \simeq 0.0072$
and the correlation parameter $r \simeq 0.99988$
(the finest division was $N=2^{25}$).

 Finally, I would like to mention that the modified scaling exponents
were suggested  to investigate QM systems with discrete spectra
\cite{StasEPSAPS,StasAPP}.
Namely, the probabilities $p_i(N)$ obtained by the simple
level--counting in a box have been changed
by summation of the de--excitation probabilities.
In this case, not only the information contained in the energy
spectrum but also in eigenfunctions is employed.
However, even in this case, up to now, we are unable to find
any interesting physical interpretation of the scaling exponents.

\acknowledgments

This research was supported by the Polish KBN Grant no 2 P03B 140 10.
The author is indebted to Professor S. Dro\.zd\.z for numerous
discussions.


\begin{references}


\bibitem{Mandelbrot} B.B. Mandelbrot, {\em The Fractal Geometry of
Nature} (Freeman, New York, 1983).

\bibitem{StasEPSAPS} A.Z. G\'orski, R. Botet, S. Dro\.zd\.z,
M. P\l oszajczak, Proceedings of the 8th Joint EPS--APS Int. Conf. on
Physics Computing, Sept 1996, eds. P. Borchers {\em et al},
pp. 37--40.

\bibitem{Saiz} A. S\'aiz, V.I. Mart\'inez, Phys. Rev. E {\bf 54},
2431 (1996).

\bibitem{StasAPP} A.Z. G\'orski, S. Dro\.zd\.z, Acta Phys. Pol.
B {\bf 28}, 1111 (1997).

\bibitem{Hofstadter} D.R. Hofstadter, Phys. Rev. B {\bf 14},
2239 (1976).

\bibitem{GeiselPHYSICA} R. Fleischmann, T. Geisel, R. Ketzmerick,
G. Petschel, Physica D {\bf 86}, 171 (1995).

\bibitem{localization} M. Kohmoto, L.P. Kadanoff, Ch. Tand, Phys.
Rev. Lett. {\bf 50}, 1870 (1983).

\bibitem{RenyiA} J. Balatony, A. Renyi, Publ. Math. Inst. Hung.
Acad. Sci. {\bf 1}, 9 (1956)  (in Hungarian) [translation:
{\em Selected papers of A. Renyi}, Vol. 1 (Budapest Academy, 1976)
p. 558].

\bibitem{RenyiB} A. Renyi, {\em Probability Theory} (North--Holland,
Amsterdam, 1970) (appendix).

\bibitem{Wigner} T.A. Brody {\em et al}, Rev. Mod. Phys. {\bf 53},
385 (1981).

\bibitem{Haake} F. Haake, {\em Quantum Signatures of Chaos}
(Springer, Berlin, 1991).

\bibitem{GrassbergerHausDim} P. Grassberger, Phys. Lett. A {\bf
107}, 101 (1985).

\bibitem{Falconer} K.J. Falconer, {\em The geometry of fractal sets},
(Cambridge University, Cambridge, 1985).

\bibitem{Molteno} T.C.A. Molteno, Phys. Rev. E {\bf 48},
R3263 (1993).


\end{references}
\end{document}